\documentclass[12pt]{article}

\usepackage{amssymb}
\usepackage{amsmath}
\usepackage{amscd}
\usepackage{latexsym}
\usepackage{graphicx}

\newcommand{\be}{\begin{equation}}
\newcommand{\ee}{\end{equation}}
\newcommand{\Dlt}{\Delta}
\newcommand{\dlt}{\delta}

\newcommand{\br}{{\bf r}}
\newcommand{\bk}{{\bf k}}

\newcommand{\bp}{{\bf p}}

\newcommand{\bt}{\beta}
\newcommand{\vp}{\varphi}

\newcommand{\al}{\alpha}

\newcommand{\sgm}{\sigma}

\newcommand{\om}{\omega}
\newcommand{\Om}{\Omega}

\newcommand{\dgr}{\dagger}
\newcommand{\lbd}{\lambda}

\newcommand{\cD}{{\cal D}}

\newcommand{\cH}{{\cal H}}

\newcommand{\rgl}{\rangle}
\newcommand{\lgl}{\langle}

\begin{document}

\begin{center}

{\Large{\bf Statistical theory of materials with nanoscale phase separation} \\ [5mm]

V.I. Yukalov$^{1*}$ and E.P. Yukalova$^{2}$ } \\ [3mm]

{\it
$^1$Bogolubov Laboratory of Theoretical Physics, \\
Joint Institute for Nuclear Research, Dubna 141980, Russia \\ [3mm]

$^2$Laboratory of Information Technologies, \\
Joint Institute for Nuclear Research, Dubna 141980, Russia }

\end{center}

\vskip 0.5cm

\begin{abstract}

Materials with nanoscale phase separation are considered. These materials are 
formed by a mixture of several phases, so that inside one phase there exist nanosize 
inclusions of other phases, with random shapes and random spatial locations. A 
general approach is described for treating statistical properties of such materials 
with nanoscale phase separation. Averaging over the random phase configurations, 
it is possible to reduce the problem to the set of homogeneous phase replicas, 
with additional equations defining the geometric weights of different phases in the 
mixture. The averaging over phase configurations is mathematically realized as the 
functional integration over the manifold indicator functions. This procedure leads 
to the definition of an effective renormalized Hamiltonian taking into account the 
existence of competing phases. Heterophase systems with mesoscopic phase separation 
can occur for different substances. The approach is illustrated by the model of a 
high-temperature superconductor with non-superconducting admixture and by the model 
of a ferroelectric with paraelectric random inclusions. 

\end{abstract}

\vskip 0.5cm

{\parindent=0pt
{\bf Keywords}: Nanoscale phase separation, High-temperature heterogeneous superconductor,
          Ferroelectric with paraelectric admixture

\vskip 0.5cm

$^*${\bf Corresponding author}: V.I. Yukalov

E-mail: yukalov@theor.jinr.ru

Phone: 7(496)216-3947 }

\vskip 1cm

\section{Introduction}

Many materials display heterogeneous phase structure, when inside a sample with 
one thermodynamic phase there occur inclusions of another phase. The known examples 
of such materials are high-temperature superconductors, where superconducting phase 
can coexist with non-superconducting phase [1-5], or ferroelectrics, where inside
ferroelectric phase there can exist the germs of paraelectric phase [6-14]. A number
of other examples can be found in the review articles [15,16]. Such materials with
heterogeneous phase structure are called heterophase. Since the inclusions of the 
other phases are usually of nanosizes, being in between the microscopic mean 
interatomic distance and the macroscopic size of the whole sample, they are termed
nanoscale or mesoscopic germs. The occurrence of such germs in a sample is called
nanoscale or mesoscopic phase separation. 

In the present paper, we consider such heterophase materials. We concentrate on the
situation, when the inclusions of the competing phases do not form regular spatial 
structures but are randomly distributed in space and have random shapes. The germs
are of nanosizes and are mesoscopic, which means that the typical size $l_g$ of
such germs is much larger than the mean interatomic distance $a$, but essentially 
smaller than the system length $L$, so that $a \ll l_g \ll L$. The germs are not 
necessarily compact, but may have various shapes. The typical size $l_g$ should be 
understood as the average size, since in reality the germs may possess a rather wide
distribution of nanoscopic sizes, being in that sense multiscale. 

In Sec. 2, we suggest a general approach for treating such materials with nanoscale
phase separation. The approach can be applied to different matters. Its application
is given in Sec. 3 for heterophase ferroelectrics and in Sec. 4, for heterophase 
superconductors. The heterophase structure of materials makes their properties much
richer than would be the properties of pure-phase systems. Section 5 is a brief 
conclusion.

\section{General approach}

Let us consider a sample that is spatially separated into two different thermodynamic 
phases, whose regions are randomly intermixed in the space. The most general way of
characterizing any kind of such a separation is by means of the Gibbs equimolecular 
separating surface [17,18]. Then the total volume $V$ of the sample is the sum of the
total volumes of the particular phases, and the total number of particles $N$ is 
also the corresponding sum:
\be
\label{1}  
 V = V_1 + V_2 \; , \qquad N = N_1 + N_2 \;  .
\ee
The snap-shot spatial locations of the phase regions are described by the manifold 
indicator function 
\begin{eqnarray}
\xi_\nu(\br) = \left \{ \begin{array}{ll}
1 , & ~ \br \in \mathbb{V}_\nu \\
0 , & ~ \br \not\in \mathbb{V}_\nu
\end{array} \; . \right.
\end{eqnarray}
The notation $V_\nu \equiv {\rm mes}{\mathbb V}_\nu$ is used. The statistical ensemble 
of the phase separated system is the pair $\{ {\cal H}, {\hat \rho(\xi)} \}$, with 
the fiber space of microstates
\be
\label{3}
 \cH = \cH_1 \bigotimes \cH_2 \; ,
\ee
being the tensor product of weighted Hilbert spaces [15,16]. The statistical 
operator ${\hat \rho}(\xi)$ satisfies the normalization condition
\be
\label{4}
{\rm Tr} \int \hat\rho(\xi)\;\cD\xi = 1\;  ,
\ee
where the trace is over the quantum degrees of freedom and the functional 
integration is over the topological space $\{ \xi \}$ of all manifold indicator 
functions. The integration over the manifold indicator functions implies the 
averaging over random phase configurations. 

To define the form of the statistical operator, we need to deal with a 
representative statistical ensemble uniquely characterizing the considered
statistical system. For this purpose, we recollect the definition of the
internal energy
\be
\label{5}
E = {\rm Tr} \int \hat\rho(\xi) \hat H(\xi) \;\cD\xi 
\ee
that is the average of the energy Hamiltonian ${\hat H}(\xi)$ . Similarly, 
there can exist the values
\be
\label{6}
C_i = {\rm Tr} \int \hat\rho(\xi) \hat C_i(\xi) \;\cD\xi
\ee
that are the averages of some constraint operators ${\hat C}_i(\xi)$. As an 
example of the latter, one can keep in mind the number-of-particle operator
associated with the conservation of the average total number of particles.  

With the given constraints, we construct the information functional
$$
I[\; \hat\rho(\xi)\; ] = {\rm Tr} \int \hat\rho(\xi) \ln\hat\rho(\xi) \;\cD\xi
\; + \; \lbd_0 \left [ {\rm Tr} \int \hat\rho(\xi) \;\cD\xi \; - 
\; 1 \right ] \; +
$$
\be
\label{7}
+ \;  \bt \left [ {\rm Tr} \int \hat\rho(\xi)\hat H(\xi) \;\cD \xi\; -
 \; E \right ] \; + \; 
\sum_i \lbd_i \left [ {\rm Tr} \int \hat\rho(\xi) \hat C_i(\xi)\;\cD\xi \; - 
\; C_i \right ] \;  ,
\ee
in which the first term is the Shannon information and $\lambda_0, \beta$, and 
$\lambda_i$ are the Lagrange multipliers guaranteeing the validity of conditions 
(4) to (6). The statistical operator is obtained from the principle of minimal 
information, that is, from minimizing the information functional, which yields
\be
\label{8}
   \hat\rho(\xi) = \frac{1}{Z} \; \exp\{ - \bt H(\xi) \} \; ,
\ee  
with the grand Hamiltonian
\be
\label{9}
H(\xi) = \hat H(\xi) - \sum_i \mu_i \hat C_i(\xi)
\ee
and partition function
\be
\label{10}
 Z = {\rm Tr} \int \exp\{ - \bt H(\xi) \} \; \cD\xi \; ,
\ee
where $\beta \equiv 1/T$ is inverse temperature and $\mu_i \equiv - \lambda_i T$. 

Accomplishing the functional integration over the manifold characteristic functions
gives the {\it effective renormalized Hamiltonian} ${\tilde H}$ defined by the
equality
\be
\label{11}
 \exp\{ - \bt \widetilde H \}  = \int \exp\{ - \bt H(\xi) \} \; \cD\xi \; .
\ee
Then the partition function (10) reduces to 
\be
\label{12}
Z = {\rm Tr} \exp\{ - \bt \widetilde H \}  ,
\ee
with the trace over the quantum degrees of freedom.

The values
\be
\label{13}
 w_\nu = \int \xi_\nu(\br) \; \cD \xi \qquad ( \nu = 1,2 ) \; ,
\ee
enjoying the properties
\be
\label{14}
 w_1 + w_2 = 1 \; , \qquad 0 \leq w_\nu \leq 1 \;  ,
\ee
are the geometric probabilities of the phases, serving as the minimizers of
the grand potential 
\be
\label{15}
 \Om = - T \ln   {\rm Tr} e^{- \bt \widetilde H}
\ee
under the normalization condition (14).  

In this way, for a considered heterophase substance, we need to find the 
effective Hamiltonian ${\tilde H}$, using relation (11), after which the 
following calculations deal with the standard methods of statistical physics,
as in expression (15) for the grand potential.

\section{Heterophase ferroelectrics}

As the first illustration of the approach, let us consider ferroelectrics 
with hydrogen bonds [10,19], such as HCl and DCl. The Hamiltonian for this 
material can be obtained starting from the usual Hamiltonian of charged particles 
in a lattice of double wells [19,20]. For hydrogen-bonds ferroelectrics the 
charged particles are protons. The scheme of the derivation is as follows. The 
field operators of the particles are expanded over Wannier functions as
\be
\label{16}
 \psi(\br) = \sum_{nj} c_{nj} \vp_{nj}(\br) \;  ,
\ee
where the multi-index $n$ labels energy bands with energies $E_n$ and the 
index $j = 1,2, \ldots$ enumerates lattice sites. The material is insulating, 
with well localized Wannier functions, so that particles do not jump between 
the lattice sites,
\be
\label{17}
 c_{mi}^\dgr c_{nj} = \dlt_{ij} c_{mj}^\dgr c_{nj} \;  .
\ee
The unity filling factor is assumed, leading to the unipolarity condition
\be
\label{18}
  \sum_{n} c_{nj}^\dgr c_{nj} = 1 \; , \qquad  c_{mj} c_{nj} = 0 \; .
\ee
Taking into account two lowest bands allows us to invoke the quasispin 
representation 
\be
\label{19}
 c_{1j}^\dgr c_{1j} = \frac{1}{2} + S_j^x \; , \qquad 
c_{2j}^\dgr c_{2j} = \frac{1}{2} - S_j^x \; , \qquad
c_{1j}^\dgr c_{2j} = S_j^z - i S_j^y
\ee
with the operators $S_j^\alpha$ satisfying the spin-half algebra. 

Following the general scheme of Sec. 2, we average over phase configurations 
obtaining the effective Hamiltonian 
\be
\label{20}
 \widetilde H = H_1 \bigoplus H_2 \;  ,
\ee
in which the Hamiltonian replicas are
\be
\label{21} 
 H_\nu = w_\nu N E_0 + \frac{w_\nu^2}{2} \sum_{i\neq j} A_{ij} - 
w_\nu \Om_0 \sum_j S_j^x + w_\nu^2 \sum_{i\neq j} \left ( B_{ij} S_i^x S_j^x - 
I_{ij} S_i^z S_j^z \right ) \;  .
\ee
Here $E_0 \equiv (E_1 + E_2)/2$, the tunneling frequency $\Omega_0 \equiv E_2 - E_1$,
and $A_{ij}, B_{ij}, I_{ij}$ are the matrix elements of the pair interaction 
potential of particles [19,20]. The phase weights $w_\nu$ are defined by minimizing
the grand potential (15).   

The phases are distinguished by the electric polarization, so that ferroelectric 
possesses a finite polarization and paralectric, zero polarization. The 
polarization is proportional to the mean imbalance
\be
\label{22}
  s_\nu \equiv \frac{2}{N} \sum_{j=1}^N \; \lgl S_j^z \rgl_\nu \; ,
\ee
which, hence, serves as an order parameter. The ferroelectric phase, with $\nu = 1$, 
is characterized by a nonzero order parameter
\be
\label{23}
 s_1\neq 0 \;  , 
\ee
while the paraelectric phase, with $\nu = 2$, by the zero order parameter
\be
\label{24}
  s_2 \equiv 0 \; .
\ee

We calculated the ferroelectric order parameter (23) in the mean-field approximation,
keeping in mind that $\Omega_0$ and $B_{ij}$ are small as compared to $I_{ij}$. The
behavior of the order parameter (23) essentially depends on the quantity
\be
\label{25}
u \equiv \frac{A}{I} \qquad \qquad 
\left ( A \equiv \frac{1}{N} \sum_{i\neq j} A_{ij} \; , ~~
   I \equiv \frac{1}{N} \sum_{i\neq j} I_{ij} \right )
\ee
characterizing the ratio of the repulsive-interaction intensity to that of attractive
interactions. Repulsive interactions oppose the system ordering, while the 
attractive interactions are responsible for this ordering. Therefore quantity (25)
plays the role of a disorder parameter. 

The behavior of the ferroelectric order parameter (23), as a function of temperature, 
is shown in Fig. 1. Temperature is measured in units of $I$. For negative values of 
$u$, phase separation cannot occur, and the sample is pure, with a second-order phase 
transition at the critical temperature $T_c = 0.5$. Nanoscale phase separation 
occurs for $u > 0$. When $0 < u < 3/2$, the phase transition to the paraelectric phase 
is of first order, happening in the range of temperatures
$$
 \frac{1}{8} \leq T_0 \leq \frac{1}{2} \;  .
$$
For $u > 3/2$ the phase transition is of second order, occurring at the tricritical 
point $T_c^* = 1/8$. As is seen, the appearance of the heterophase states makes
the phase-transition picture essentially richer that it would be for a pure 
ferroelectric phase, where there can happen only the second-order phase transition 
at $T_c = 0.5$.

\section{Heterophase superconductors}

Mesoscopic phase separation in high-temperature superconductors has been documented
in many experiments [1-5,21-23]. Yet before the high-temperature superconductivity 
was discovered [24], there had been suggestions [25,26] for the possibility of 
nanoscopic phase separation in superconductors. These and the following theoretical 
models [25-31] have shown that the arising heterophase states in superconductors are 
quite natural, rendering the material more stable, as compared to the pure-phase 
sample.

The origin of high-temperature superconductivity has not yet been unambiguously 
clarified. There exist several mechanisms attempting to explain this phenomenon.
The aim of the present paper is not to compare different models of describing 
high-temperature superconductivity or to study their relation to experiment, but
our goal is to analyze, in the frame of one model, based on one chosen approximation,
how the appearance of nanoscale phase separation changes the properties of the system,
as compared to the same approximation, but without taking into account the phase 
separation. For this purpose, we start with the standard Hamiltonian of charged 
particles, whose field operators are expanded over the basis of Bloch functions
\be
\label{26}
  \psi_s(\br) = \sum_k c_s(\bk) \vp_k(\br) \; ,
\ee
where the index $s$ denotes spin, while $\bf k$ is quasimomentum. The problem can 
be simplified by involving the Bardeen-Cooper-Schrieffer restriction
\be
\label{27}
 c_s^\dgr(\bk) c_r(\bp) = \dlt_{sr}\dlt_{kp}  c_s^\dgr(\bk) c_s(\bk) \; , 
\qquad 
c_s(\bk) c_r(\bp) = \dlt_{-sr}\dlt_{-kp}  c_s(\bk) c_{-s}(-\bk) \; .
\ee

Assuming the occurrence of nanoscale phase separation and averaging over phase 
configurations, according to Sec. 2, we come to the effective Hamiltonian (20), 
with the Hamiltonian replicas
$$
H_\nu = w_\nu \sum_s \; \sum_{kp} \; [\; t_\nu(\bk,\bp) - 
\mu\dlt_{kp} \; ] \; c_s^\dgr(\bk) c_s(\bp) \; +
$$
\be
\label{28}
 + \; \frac{w_\nu^2}{2} \; \sum_{ss'} \; \sum_{kk'} \; \sum_{pp'}
V_\nu(\bk,\bk',\bp',\bp)\; c_s^\dgr(\bk)c_{s'}(\bk')  c_{s'}(\bp')c_s(\bp) \; ,
\ee
defined on the space ${\cal H}_\nu$, and in which $t_\nu({\bf k}, {\bf p})$ is 
the transport matrix, $\mu$ is chemical potential, and 
$V_\nu({\bf k},{\bf k^\prime},{\bf p^\prime},{\bf p})$ is the interaction vertex.

The phases are distinguished by the order parameters that are given either by the
anomalous averages
\be
\label{29}
\sgm_\nu(\bk) \equiv \lgl c_{-s}(-\bk) c_{s}(\bk) \rgl_\nu \; ,
\ee
or by the related gaps
\be
\label{30}
 \Dlt_\nu(\bk) = w_\nu \sum_p J_\nu (\bk,\bp) \sgm_\nu(\bp) \;  ,
\ee
where $\nu = 1,2$. Statistical averages for the superconducting phase are defined 
in the space ${\cal H}_1$ with broken gauge symmetry, because of which  
\be
\label{31}
 \sgm_1(\bk) \not\equiv 0 \; , \qquad   \Dlt_1(\bk) \not\equiv 0 \;,
\ee
while for the normal phase, gauge symmetry of microstates in ${\cal H}_2$ is 
preserved, yielding
\be
\label{32}
  \sgm_2(\bk) \equiv 0 \; , \qquad   \Dlt_2(\bk) \equiv 0 \;  .
\ee
 
Next, we use the Hartree-Fock-Bogolubov approximation, which, together with the 
Bardeen-Cooper-Schrieffer restriction (27), gives the superconducting gap equation 
\be
\label{33}
 \Dlt_1(\bk) = \frac{w_1^2}{2} \; \sum_p J_1(\bk,\bp) \; 
\frac{\Dlt_1(\bp)}{E_1(\bp)} \; \tanh \; \frac{w_1 E_1(\bp)}{2T} \; ,
\ee
in which $E_1({\bf p})$ is the Bogolubov spectrum of excitations. The similar
equation for the normal phase is trivial, giving zero gap, according to 
condition (32). As is seen, the equation contains the phase probability $w_1$,
characterizing the geometric weight of the superconducting phase. In order that
the heterophase superconductor be stable, the grand potential (15) is to be 
minimal with respect to this weight. The positivity of the second derivative
of the grand potential plays the role of stability condition and requires the 
existence of effective Coulomb interactions $Q > 0$. That is: {\it nanoscale 
phase separation in superconductors can be thermodynamically stable only in the 
presence of repulsive Coulomb interactions}. 

Superconductivity, as such, presupposes that the gap is not identically zero, 
in agreement with definition (31). This leads to the superconductivity criterion
\be
\label{34}
 \frac{|\al|^2}{\om_0^2} \; - \; w_1 Q > 0 \;  ,
\ee
where $\alpha$ is the charge-lattice coupling and $\omega_0$ is the 
characteristic lattice frequency.

In a pure-phase sample, where $w_1 = 1$, strong Coulomb interactions may hamper
the existence of superconductivity, increasing the second term of inequality (34). 
But the same sample can become superconducting, if phase separation occurs, so that
$w_1$ becomes less than one, thus diminishing the second term in criterion (34). 
Therefore: {\it nanoscale phase separation enables the appearance of superconductivity
even if it were impossible in pure-phase matter}.

Introducing the dimensionless parameter $\mu^*$, the superconductivity criterion
(34) can be written in the form
\be
\label{35}
 1 - w_1\mu^* > 0 \qquad \left ( \mu^* \equiv \frac{\om_0^2}{|\al|^2} \; Q
\right ) \;  .
\ee
In the case of good conductors, $\mu^* \ll 1$, and the criterion is easily satisfied. 
However, for bad conductors, $\mu^* > 1$, so that criterion (35) can be broken in a 
pure sample, where $w_1 = 1$. But, in the presence of phase separation, when 
$w_1 < 1$, this criterion can again be valid. Hence: {\it nanoscale phase separation 
facilitates the occurrence of superconductivity in bad conductors}.           

Keeping in mind cuprates [32-35], let us consider superconductivity in a 
two-dimensional plane, with a square lattice of lattice spacing $a$. We shall 
use the dimensionless quasimomenta ${\bf k} \equiv {\bf p} a$. The gap in 
high-temperature superconductors is usually anisotropic and can be represented 
as an expansion 
$$
\Dlt_1(\bk) = \sum_i \Dlt_i\chi_i(\bk)
$$
over a basis of functions characterizing the existing lattice symmetry. For a square 
lattice, one has the functions 
\begin{eqnarray}
\nonumber
\begin{array}{ll}
\chi_1(\bk) = 1 & ~~~ ( s \; symmetry) \; ,\\
\\
\chi_2(\bk) = \frac{1}{2}\; (\cos k_1 + \cos k_2) & ~~~ ( s^* \; symmetry)\; , \\
\\
\chi_3(\bk) = \cos k_1 \cdot \cos k_2 & ~~~ ( s_{xy} \; symmetry) \; , \\
\\
\chi_4(\bk) = \sin k_1 \cdot \sin k_2 & ~~~ ( d_{xy} \; symmetry) \; , \\
\\
\chi_5(\bk) = \frac{1}{2}\; ( \cos k_1 - \cos k_2 ) & ~~~ ( d_{x^2-y^2} \; symmetry) \; .
\end{array}
\end{eqnarray}
The prevalence of this or that symmetry is prescribed by the largest density of states.
Accepting the parameters typical of cuprates, we come to the conclusion that the largest
density of states occurs for the mixture of $s$ and $d_{x^2 - y^2}$ symmetries. 

Finally, assuming that the superconducting weight is proportional to the doping, we
calculate the superconducting transition temperature $T_c$, which is presented in
Fig. 2. As we see: {\it nanoscale phase separation results in a bell shape of the 
superconducting temperature as a function of the superconducting phase weight}.

\section{Conclusion}

Nanoscale phase separation can occur in different types of matter. When the phase 
separation arises in the form of nanoscale germs of competing phases, randomly 
distributed in space, then it is possible to introduce the averaging over the random 
phase configurations and reduce the consideration to the study of an effective
Hamiltonian renormalized with phase weights that serve as the minimizers of the 
grand potential. Therefore the phase separated sample can be more stable than the 
pure-phase one. We illustrated the general approach by two models taking into account
the phase separation. The first model is a ferroelectric with the germs of 
paraelectric phase. The second example is a superconductor with the germs of normal
phase. The occurrence of the nanoscale phase separation in materials essentially
changes their properties, yielding a number of effects that would be absent in 
pure-phase samples, without phase separation.     

\vskip 5mm

The authors declare that they have no conflict of interest.

\vskip 1cm

\begin{figure}[ht]
\centerline{
\includegraphics[width=8cm]{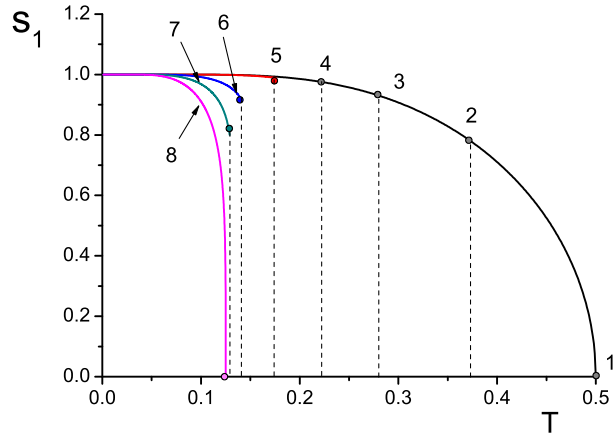}}
\caption{Ferroelectric order parameter as a function of temperature 
(in units of $I$) for different disorder parameters: (1) $u = 0$; (2) $u = 0.1$; 
(3) $u = 0.3$; (4) $u = 0.4$; (5) $u = 0.51$; (6) $u = 0.75$; (7) $u = 1$; 
(8) $u = 1.5$. The points mark the temperatures of ferroelectric-paraelectric 
phase transitions.}
\label{fig:Fig.1}
\end{figure}

\vskip 1cm

\begin{figure}[ht]
\centerline{
\includegraphics[width=8cm]{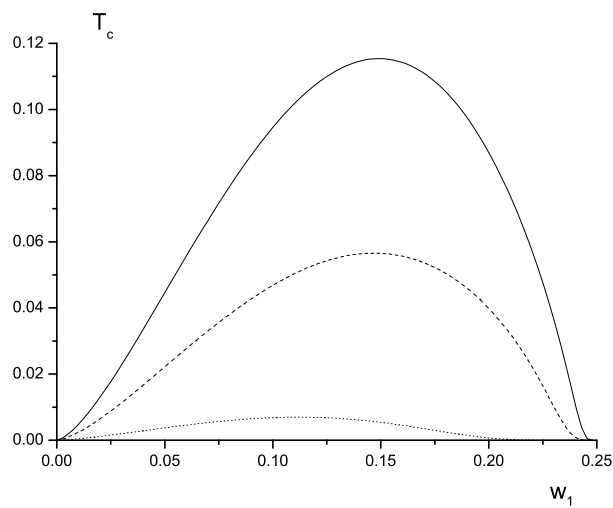} }
\caption{Critical temperature $T_c$ (in units of $\omega_0$) as a function 
of the superconducting weight $w_1$ for different effective couplings: 
$\lambda = 1$ (dotted line); $\lambda = 5$ (dashed line); $\lambda = 10$ 
(solid line).}
\label{fig:Fig.2}
\end{figure}

\end{document}